\documentclass[aps,prb,reprint,superscriptaddress,showpacs,floatfix,nofootinbib,longbibliography]{revtex4-2}

\usepackage{makecell}
\usepackage{tabularx}
\usepackage{array}
\newcolumntype{Y}{>{\raggedright\arraybackslash}X}
\usepackage{amsmath}
\usepackage{amssymb}
\usepackage{graphicx}
\usepackage{bm}
\usepackage{hyperref}
\usepackage{siunitx}
\usepackage{color}
\usepackage{booktabs}
\hypersetup{
  colorlinks=true,
  linkcolor=blue,
  citecolor=blue,
  urlcolor=blue
}

\newcommand{\Tc}{T_{\mathrm{c}}}

\begin{document}

\title{
A Modulated Electron Lattice (MEL) Criterion for Metallic Superconductivity
}

\author{Jaehwahn Kim}
\affiliation{Hyunsung T\&C Laboratory, Suwon 16679, Republic of Korea}

\author{Davis A.~Rens}
\affiliation{Department of Physics, University of California, Berkeley, California 94720, USA}

\author{Waqas Khalid}
\affiliation{Department of Physics, University of California, Berkeley, California 94720, USA}

\author{Hyunchul Kim}
\affiliation{Hyunsung T\&C Laboratory, Suwon 16679, Republic of Korea}

\date{\today}

\begin{abstract}
A central unresolved question in the theory of superconductivity is why only a small subset 
of metallic elements exhibit a superconducting state, whereas many others remain strictly normal.  
Neither the conventional Bardeen Cooper Schrieffer (BCS) framework nor its extensions involving 
charge density wave (CDW) or pair density wave (PDW) order provide a predictive or 
material-selective criterion capable of distinguishing superconducting metals from 
non-superconducting ones.  
In particular, the persistent absence of superconductivity in simple noble metals with 
well-defined Fermi surfaces poses a challenge for all traditional approaches.

Here we address this problem using the Modulated Electron Lattice (MEL) 
Ginzburg Landau (GL) framework introduced in our previous work.  
In this formulation, a coarse-grained MEL charge field $\rho_{\mathrm{MEL}}(\mathbf{r})$ with 
momentum dependent stiffness $\alpha(q)$ is coupled to the superconducting (SC) order parameter 
$\psi(\mathbf{r})$.  
We show that metallic superconductivity emerges only when the system satisfies a 
specific ``MEL enhancement window,'' characterized by a negative minimum of $\alpha(q)$ at either 
a finite modulation wave vector $q^{\ast}$ or at $q=0$, together with sufficiently strong 
coupling between $\rho_{\mathrm{MEL}}$ and $\psi$.

This unified criterion naturally partitions metallic elements into three universal classes:  
(i) MEL-enhanced superconductors with a finite-$q^{\ast}$ charge mode,  
(ii) conventional BCS superconductors as the homogeneous $q^{\ast}=0$ limit of the MEL framework,  
and (iii) metals for which $\alpha(q)$ remains positive for all $q$, suppressing all MEL modes and
preventing any superconducting instability.  

By applying this criterion to simple metallic elements, we identify why some metals develop superconductivity while others do not, possibly resolving a selection problem that remained open for decades 
within the BCS paradigm.  
\end{abstract}
\maketitle

\section{Introduction}

A long-standing challenge in the theory of superconductivity is to understand why only a limited 
subset of metallic elements become superconducting while many others remain strictly normal, 
even at the lowest experimentally accessible temperatures.  
Within the conventional Bardeen Cooper Schrieffer (BCS) framework 
\citep{bardeen1957theory,schrieffer1983theory}, superconductivity arises from a 
phonon-mediated effective attraction that drives Cooper pairing.  
Although this paradigm successfully explains numerous thermodynamic and spectroscopic properties 
of classic low-$T_{\mathrm{c}}$ superconductors such as Al, Sn, and Pb, it does not provide a 
predictive principle capable of distinguishing superconducting metals from non-superconducting ones.  
Key quantities such as the electron--phonon coupling constant, the Coulomb pseudopotential, and the 
characteristic phonon energy enter as phenomenological inputs, limiting the framework's ability to 
explain the empirical absence of superconductivity in simple metals like Cu, Ag, and Au.

Extensions of the BCS theory incorporating charge density wave (CDW) or pair density wave (PDW) 
order \citep{gruner1988dynamics,fradkin2015colloquium,agterberg2020physics} have shed light on 
competing or intertwined ordering phenomena, particularly in correlated systems such as the cuprates.  
However, these approaches also lack a general, material-selective criterion for identifying 
which metals can host a superconducting state.  
The characteristic ordering wave vector in CDW or PDW theories is typically inferred from 
experimental observations or specific Fermi surface geometries, rather than emerging from a universal principle applicable across elemental metals.

In this work we present a unified explanation of metallic superconductivity using the 
Modulated Electron Lattice (MEL) Ginzburg Landau (GL) framework introduced in our previous study 
\citep{kim2025mel}.  
In this formulation, a coarse-grained charge modulation field $\rho_{\mathrm{MEL}}(\mathbf{r})$ 
is coupled to the superconducting (SC) order parameter $\psi(\mathbf{r})$ through a 
momentum-dependent stiffness $\alpha(q)$ and well-defined MEL--SC coupling terms.  
The resulting free energy functional yields a material-selective criterion for superconductivity 
based on the presence or absence of a ``MEL enhancement window'' — a range of momenta for which 
$\alpha(q)$ becomes negative and stabilizes either a homogeneous ($q=0$) or modulated ($q^{\ast}$) 
charge mode capable of enhancing the SC sector.

This framework naturally classifies metallic elements into three universal regimes:  
(i) systems with a finite-$q^{\ast}$ charge mode that amplifies the effective SC stiffness 
(MEL enhanced superconductors),  
(ii) systems in which only the homogeneous $q=0$ charge mode is stabilized 
(conventional BCS superconductors),  
and (iii) systems in which $\alpha(q)$ remains strictly positive for all momenta, thereby 
suppressing all MEL modes and preventing superconductivity.  

In the remainder of this paper, we establish these criteria rigorously, evaluate their 
consequences for representative metallic elements, and demonstrate how this unified framework 
accounts for both superconducting and non-superconducting metals within a single theoretical 
structure.

\section*{Notation and Abbreviations}

We summarize in Table \ref{abbrev} the key symbols, fields, and abbreviations used throughout this work.

\begin{table}[t]
\centering
\begin{tabular}{ll}
\label{abbrev}
Symbol & Definition \\
\hline
MEL & Modulated Electron Lattice \\
GL  & Ginzburg--Landau framework \\
SC  & Superconducting phase or order parameter sector \\
CDW & Charge-Density-Wave order \\
PDW & Pair-Density-Wave order \\
$\psi(\mathbf{r})$ & Superconducting order parameter \\
$\rho_{\mathrm{MEL}}(\mathbf{r})$ & MEL charge-modulation amplitude \\
$q^{\ast}$ & Preferred MEL modulation wave vector \\
$\alpha(q)$ & Momentum-dependent MEL stiffness \\
$\chi_{\mathrm{el}}(q)$ & Electronic charge susceptibility \\
$D_{\mathrm{ph}}(q)$ & Phonon propagator contribution \\
$K_s, K_\rho$ & SC gradient stiffness and analytic $q^2$ MEL stiffness \\
$\gamma_1, \gamma_2$ & MEL--SC coupling constants \\
$\Tc$ & Superconducting critical temperature \\
\hline
\end{tabular}
\caption{Notation and abbreviations.}
\end{table}

\section{Minimal MEL Criteria for Metallic Superconductivity}

\subsection{Ginzburg--Landau formulation of the coupled MEL--SC fields}

We start from the coupled Ginzburg--Landau (GL) functional introduced in our previous work,\cite{kim2025mel}
written in terms of the superconducting (SC) order parameter $\psi(\mathbf r)$ and a coarse-grained
modulated electron lattice (MEL) charge field $\rho_{\mathrm{MEL}}(\mathbf r)$:
\begin{align}
F[\psi,\rho_{\mathrm{MEL}}]
= \int d^3 r \Big\{
& \alpha_s |\psi|^2
+ \frac{\beta_s}{2} |\psi|^4
+ K_s |\nabla \psi|^2                                      \notag \\
&+ \frac{1}{2}\rho_{\mathrm{MEL}}\,\alpha(-i\nabla)\,\rho_{\mathrm{MEL}}
+ \frac{\beta_\rho}{4} \rho_{\mathrm{MEL}}^4               \notag \\
&+ \gamma_1 \rho_{\mathrm{MEL}} |\psi|^2
+ \gamma_2 \rho_{\mathrm{MEL}}^2 |\psi|^2
\Big\}.
\label{eq:GL_total}
\end{align}

\noindent\textbf{Definitions.}
$\psi(\mathbf r)$ is a complex SC order parameter and $\rho_{\mathrm{MEL}}(\mathbf r)$ is a real MEL field.
$\alpha_s(T)$ is the bare SC quadratic coefficient (positive in a normal metal and changing sign at $T_c$ in an
intrinsic BCS superconductor), while $\beta_s>0$ and $K_s>0$ are the usual quartic and gradient coefficients.
The MEL sector is controlled by a nonlocal harmonic kernel $\alpha(-i\nabla)$, a stabilizing quartic term
$\beta_\rho>0$, and MEL--SC couplings $\gamma_{1,2}$.

\begin{figure}[t]
\centering
\includegraphics[width=0.92\linewidth]{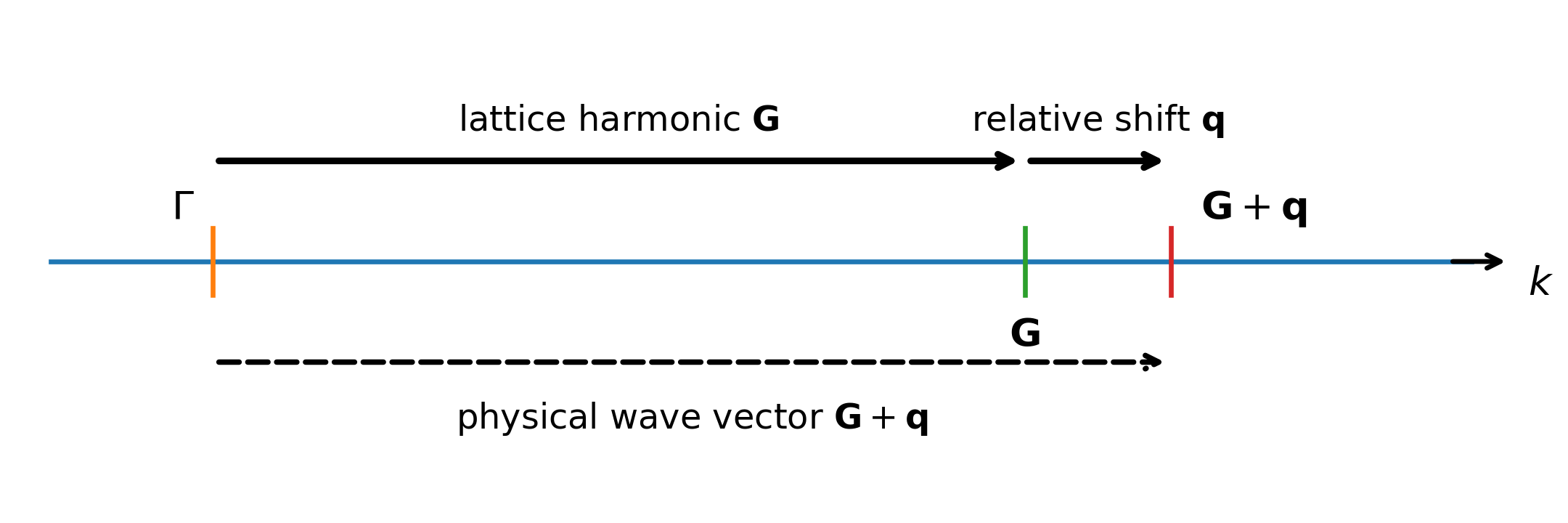}
\caption{Definition of the relative wave vector $\mathbf q$ used in the MEL Fourier expansion.  The physical modulation
wave vector is $\mathbf G+\mathbf q$, i.e., $\mathbf q$ is a small deviation measured relative to a reciprocal-lattice
harmonic $\mathbf G$.  With this convention, $q^\ast=0$ is lattice-locked (commensurate).}
\label{fig:q_relative}
\end{figure}

Here $\mathbf q$ is a \emph{relative} modulation wave vector defined with respect to an underlying reciprocal-lattice
harmonic: $\rho_{\mathrm{MEL}}$ is the slowly varying envelope of an electronic density component at $\mathbf G+\mathbf q$
(Fig.~\ref{fig:q_relative}).  Consequently, $q^\ast=0$ corresponds to a lattice-locked (commensurate) component rather
than a uniform change in the total charge density.

\subsection{Momentum-dependent MEL stiffness and the enhancement window}

In the MEL framework the harmonic stiffness $\alpha(q)$ is momentum dependent.  In the present work we
parametrize the static (low-frequency) kernel as
\begin{equation}
\alpha(q) = \alpha_0 + K_\rho q^2
+ c_{\mathrm{el}} \, \chi_{\mathrm{el}}(q)
+ c_{\mathrm{ph}} \, D_{\mathrm{ph}}(q),
\label{eq:alphaq}
\end{equation}
where $\chi_{\mathrm{el}}(q)$ is the (static) electronic charge susceptibility, $D_{\mathrm{ph}}(q)$ encodes the
relevant lattice/phonon contribution, $c_{\mathrm{el}}$ and $c_{\mathrm{ph}}$ are coupling coefficients, $\alpha_0$
is a bare harmonic term, and $K_\rho>0$ provides the analytic $q^2$ stiffness at small momenta.

We define the \emph{softest MEL mode} by
\begin{equation}
q^\ast = \arg\min_q \alpha(q),\qquad
\alpha_{\min}\equiv \alpha(q^\ast).
\label{eq:qstar_def}
\end{equation}
A material is said to lie inside the \emph{MEL enhancement window} when $\alpha_{\min}$ is sufficiently small
(in particular, when it crosses through zero at mean-field level) so that MEL amplitudes or fluctuations become
strong and can strongly renormalize the SC sector via the couplings $\gamma_{1,2}$.

\begin{figure}[t]
\centering \includegraphics[width=0.46\textwidth]{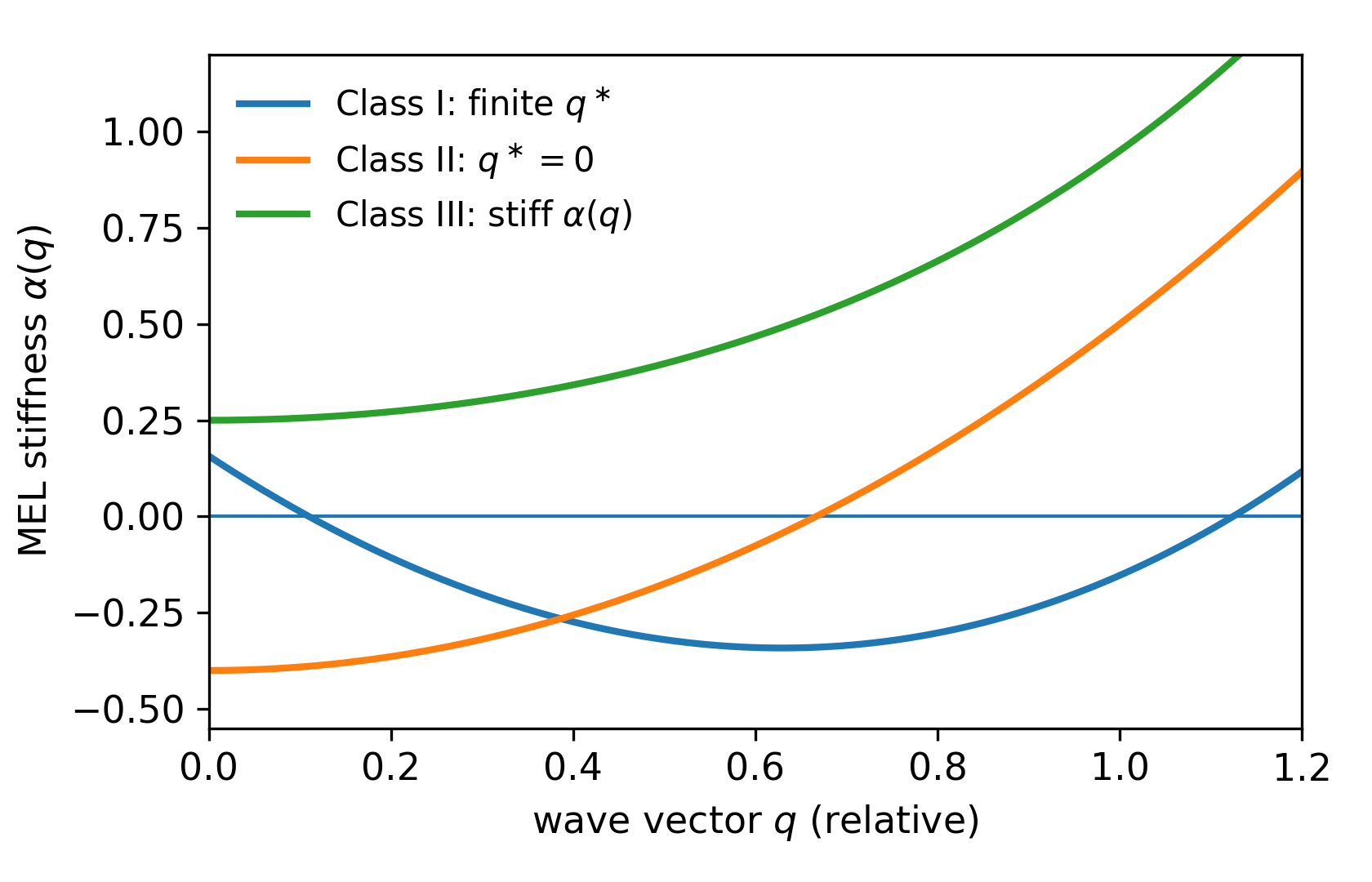}%
\caption{
Schematic momentum dependence of the MEL stiffness $\alpha(q)$ illustrating the three universal metallic
classes introduced below: Class~I (finite-$q^\ast$ soft mode), Class~II (homogeneous $q^\ast=0$ soft mode),
and Class~III (no soft MEL mode, $\alpha(q)>0$ for all $q$).
}
\label{fig:alpha_classes}
\end{figure}

\subsection{Static order versus soft-mode fluctuations}
\label{subsec:static_vs_fluct}

In a purely static mean-field GL theory, $\alpha_{\min}<0$ implies that the MEL field wants to condense at
$q^\ast$.
For $q^\ast\neq 0$ (measured relative to the lattice-locked component) this corresponds to an additional,
spatially modulated charge pattern and thus breaks lattice-translation symmetry; the $q^\ast=0$ limit is
commensurate with the underlying ionic lattice and does not introduce any further translational symmetry breaking.
However, the central quantity that enters the SC sector through Eq.~(\ref{eq:GL_total}) is not the mere
existence of long-range static order, but the \emph{MEL amplitude} as encoded in expectation values such as
$\langle\rho_{\mathrm{MEL}}\rangle$ and $\langle\rho_{\mathrm{MEL}}^2\rangle$.
In particular, $\rho_{\mathrm{MEL}}$ is a coarse-grained ``electron--lattice coherence'' field: strong
near-critical fluctuations, disorder, or anharmonicity can yield a large mean-square amplitude even when
long-range static order is not observed.

At the Gaussian level (keeping only $F_{\rho}^{(2)}$), equipartition gives a useful estimate for the
mean-square fluctuation of a mode,
\begin{equation}
\langle|\rho_{\mathbf q}|^2\rangle \simeq \frac{k_B T}{\alpha(q)} ,
\label{eq:rhoq_fluct}
\end{equation}
and therefore a schematic estimate for the local mean-square amplitude
\begin{equation}
\langle\rho_{\mathrm{MEL}}^2\rangle \sim \int\!\frac{d^3q}{(2\pi)^3}\,\frac{k_B T}{\alpha(q)} .
\label{eq:rho2_fluct}
\end{equation}

\noindent\textbf{Remark (thermal vs.\ quantum fluctuations).} Equations~(\ref{eq:rhoq_fluct})--(\ref{eq:rho2_fluct})
use a classical (equipartition) estimate for transparency.  At low temperatures the same enhancement persists with
$k_BT$ replaced by the appropriate quantum fluctuation factor, but the soft denominator $\alpha(q)$ and the phase space
around $q^\ast$ remain the controlling ingredients.

Equations~(\ref{eq:rhoq_fluct})--(\ref{eq:rho2_fluct}) also make explicit why a deep minimum in $\alpha(q)$ can
strongly enhance $\langle\rho_{\mathrm{MEL}}^2\rangle$ even in the absence of a sharp thermodynamic CDW
transition: the soft-mode phase space and the small denominator amplify the mean-square MEL amplitude.

\subsection{Coupling-induced renormalization of the SC quadratic coefficient}
\label{subsec:coupling_mass}

The MEL--SC couplings in Eq.~(\ref{eq:GL_total}) do not mix $\psi$ and $\rho_{\mathrm{MEL}}$ at quadratic order
about the trivial normal state $(\psi,\rho_{\mathrm{MEL}})=(0,0)$: the leading couplings are cubic
($\gamma_1\rho_{\mathrm{MEL}}|\psi|^2$) and quartic ($\gamma_2\rho_{\mathrm{MEL}}^2|\psi|^2$).
Consequently, a superconducting instability is not obtained from a negative eigenvalue of a \emph{quadratic}
$(\psi,\rho_{\mathrm{MEL}})$ form.  Instead, superconductivity is promoted or suppressed through the
renormalization of the effective SC ``mass'' in the presence of a finite MEL amplitude or strong MEL
fluctuations.

To leading order in the GL expansion, the effective SC quadratic coefficient is shifted by MEL expectation values:
\begin{equation}
\alpha_s^{(\mathrm{eff})}(T)
=
\alpha_s(T)
+ \gamma_1 \langle \rho_{\mathrm{MEL}} \rangle
+ \gamma_2 \langle \rho_{\mathrm{MEL}}^2 \rangle .
\label{eq:alpha_eff_general}
\end{equation}

For Class~II systems with a homogeneous ($q^\ast=0$) MEL mode, $\langle\rho_{\mathrm{MEL}}\rangle=\rho_0\neq 0$
is allowed and both $\gamma_1$ and $\gamma_2$ can contribute.
For Class~I systems with a finite-$q^\ast$ MEL mode, symmetry implies $\langle\rho_{\mathrm{MEL}}\rangle=0$ in the
absence of domain imbalance, but $\langle\rho_{\mathrm{MEL}}^2\rangle\neq 0$ once a finite-$q^\ast$ MEL amplitude
or fluctuation spectrum is present; in this case the $\gamma_2$ term provides a generic spatially averaged shift
of the SC mass term.

A superconducting transition occurs when
\begin{equation}
\alpha_s^{(\mathrm{eff})}(T_c)=0.
\label{eq:Tc_condition_general}
\end{equation}
In a metal that would remain normal in the absence of MEL effects (i.e., $\alpha_s(T)>0$ over the experimentally
accessible range), Eq.~(\ref{eq:Tc_condition_general}) shows that superconductivity requires (i) a sufficiently
soft MEL sector (large $\langle\rho_{\mathrm{MEL}}\rangle$ and/or $\langle\rho_{\mathrm{MEL}}^2\rangle$) and
(ii) sufficiently strong MEL--SC coupling of the sign that \emph{reduces} $\alpha_s^{(\mathrm{eff})}$ (e.g.,
$\gamma_2<0$ for the generic finite-$q^\ast$ mechanism).

\subsection{Unified MEL classification of metallic elements}
\label{subsec:mel_classes}

Equations~(\ref{eq:alphaq})--(\ref{eq:Tc_condition_general}) motivate a universal classification of metals based
on the structure of the MEL stiffness $\alpha(q)$:

\begin{enumerate}
\item \textbf{Class I (finite-$q^\ast$ MEL-enhanced superconductors):}
$\alpha(q)$ develops its minimum at a nonzero wave vector $q^\ast\neq 0$, yielding a soft finite-$q^\ast$ MEL mode.
Superconductivity is enhanced or induced primarily through the mean-square amplitude
$\langle\rho_{\mathrm{MEL}}^2\rangle$ entering Eq.~(\ref{eq:alpha_eff_general}) (generic $\gamma_2$ mechanism).

\item \textbf{Class II (homogeneous MEL/BCS limit):}
the softest mode is the homogeneous $q^\ast=0$ limit.  The MEL framework then reduces to the conventional GL/BCS
phenomenology in which a uniform MEL background renormalizes the pairing tendency.

\item \textbf{Class III (MEL-suppressed normal metals):}
$\alpha(q)$ remains positive and sufficiently large for all momenta.  Then MEL amplitudes and fluctuations are
small, $\alpha_s^{(\mathrm{eff})}\approx \alpha_s>0$, and no superconducting instability occurs.
\end{enumerate}

\begin{table*}[t]
\begingroup
\squeezetable             
\setlength{\tabcolsep}{4pt}            
\centering
\begin{ruledtabular}
\begin{tabular}{@{}l@{}l@{}l@{}l@{}}
Class &
Condition on $\alpha(q)$ &
Generic SC renormalization channel &
Representative examples \\
\hline
I   &
\makecell[l]{%
$\min_q \alpha(q)$ at $q^\ast\neq 0$\\
(soft finite‑$q^\ast$ mode)%
} &
\makecell[l]{%
$\alpha_s^{(\mathrm{eff})} = \alpha_s + \gamma_2\langle \rho_{\mathrm{MEL}}^2\rangle$\\
(typically $\langle \rho\rangle = 0$)%
} &
\makecell[l]{%
Metals with incipient finite‑$q$ charge/structural\\
soft modes (e.g.\ CDW‑prone systems)%
} \\[0.4em]
II  &
\makecell[l]{%
minimum at $q^\ast=0$\\
(homogeneous soft mode)%
} &
$\alpha_s^{(\mathrm{eff})} = \alpha_s + \gamma_1\rho_0 + \gamma_2\rho_0^2$ &
\makecell[l]{%
Conventional elemental superconductors\\
(Al, Sn, Pb, Nb, \ldots)%
} \\[0.4em]
III &
\makecell[l]{%
$\alpha(q)>0$ and not soft\\
for all $q$%
} &
\makecell[l]{%
No significant MEL amplitude;\\
$\alpha_s^{(\mathrm{eff})}\approx\alpha_s>0$%
} &
\makecell[l]{%
Noble metals (Cu, Ag, Au) and other\\
non‑superconducting elements at ambient pressure%
} \\
\end{tabular}
\end{ruledtabular}
\caption{Universal MEL classification of metals within the GL framework. Class I and II require a soft MEL sector (a deep minimum of $\alpha(q)$), but superconductivity additionally depends on the sign and magnitude of the MEL--SC couplings through Eq.~(\ref{eq:alpha_eff_general}).}
\label{tab:metal_classes}
\endgroup
\end{table*}

\subsection{Connection to observables and first-principles extraction of $\alpha(q)$}
\label{subsec:observables}

Equation~(\ref{eq:alphaq}) is intended as a \emph{coarse-grained} static kernel.
Microscopically, both $\chi_{\mathrm{el}}$ and $D_{\mathrm{ph}}$ arise from retarded charge and phonon response
functions; in the GL free energy we use their low-frequency (static) limit.
From a first-principles perspective, $\alpha(q)$ can be constrained or extracted using standard electronic-structure
tools such as density-functional perturbation theory (DFPT) and electron--phonon calculations.\cite{baroni2001dfpt,giustino2017eph,grimvall1981electronphonon}
Concretely, one may (i) compute phonon dispersions and electron--phonon matrix elements via DFPT,
(ii) evaluate the static electronic susceptibility $\chi_{\mathrm{el}}(q)$ (e.g., Lindhard/RPA using the DFT band
structure), and (iii) combine these ingredients to construct the effective stiffness $\alpha(q)$ and locate its
minimum $q^\ast$.
In Sec.~V we discuss how existing SCDFT benchmark calculations and electron--phonon coupling data can be used as
an empirical proxy for this program when a full $\alpha(q)$ extraction is not yet performed for a given element.

\section{Conventional BCS Metals as the \texorpdfstring{$q^{\ast}=0$}{q*=0} Limit of the MEL Framework}
In this section we demonstrate that conventional Bardeen Cooper Schrieffer (BCS) 
superconductors naturally emerge as a special case of the Modulated Electron Lattice (MEL) 
Ginzburg Landau (GL) framework when the minimum of the momentum-dependent stiffness $\alpha(q)$ 
occurs at $q^{\ast}=0$.  
This limit corresponds to the stabilization of a lattice-locked (commensurate) MEL envelope
$\rho_{\mathrm{MEL}}(\mathbf r)=\mathrm{const}$ in the MEL convention (Fig.~\ref{fig:q_relative}),
meaning that the charge sector does not favor any \emph{additional} finite-wave-vector modulation beyond the
underlying reciprocal-lattice harmonic.
\begin{figure}[t]
\centering
\includegraphics[width=0.46\textwidth]{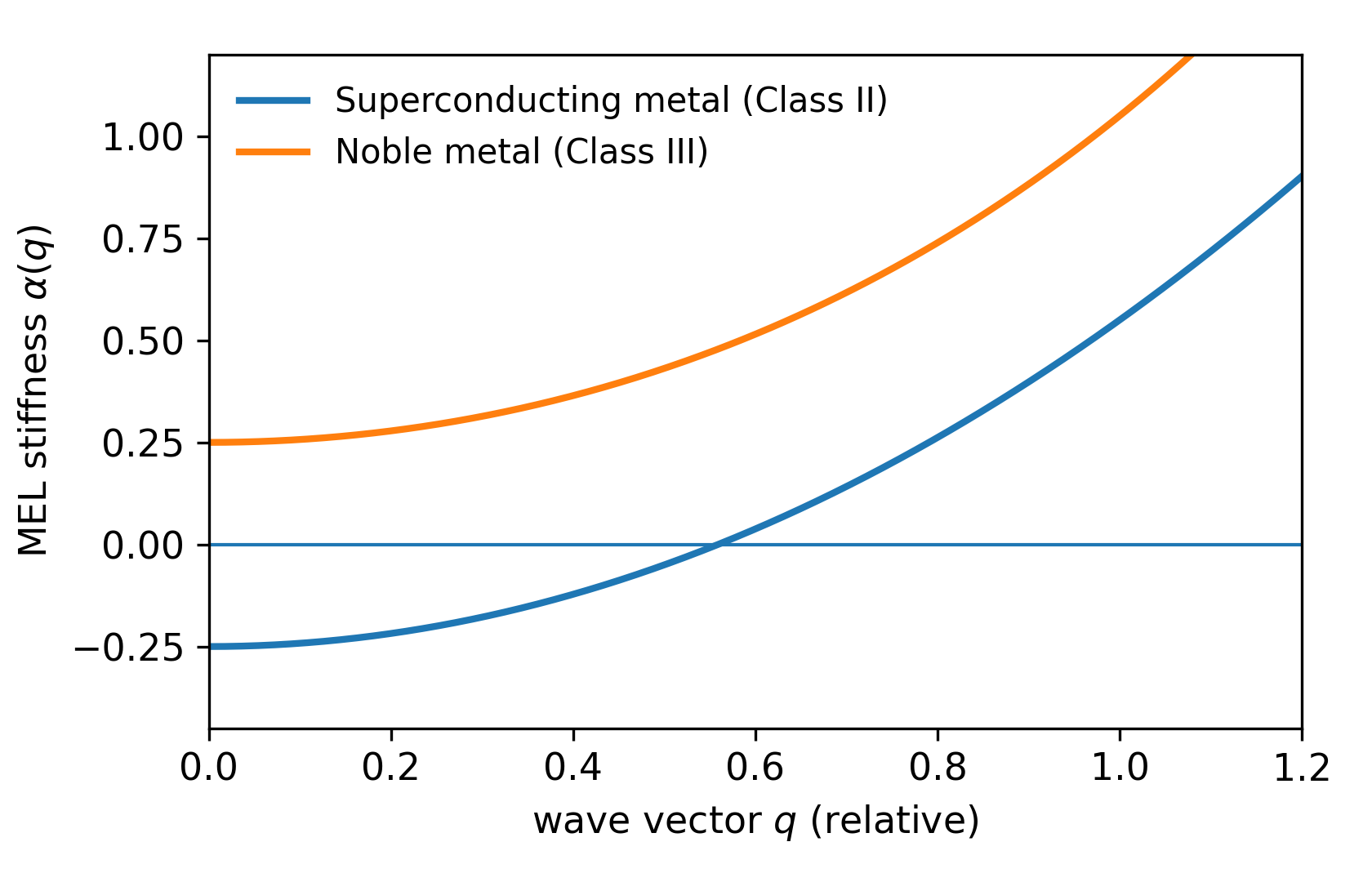}
\caption{
Schematic comparison of $\alpha(q)$ for conventional elemental superconductors (e.g., Al and Pb) and noble
metals (Cu, Ag, Au).  In the MEL language the former lie closer to a homogeneous ($q^\ast=0$) softening,
whereas the latter remain comparatively stiff with $\alpha(q)>0$ for all momenta.
}
\label{fig:noble_vs_sc}
\end{figure}

\subsection{Homogeneous MEL field and effective free energy}

When $\alpha(q)$ attains its minimum at $q=0$ and satisfies
\begin{equation}
\alpha(0) < 0, 
\label{eq:alpha_zero_negative}
\end{equation}
the MEL field develops a uniform amplitude $\rho_{\mathrm{MEL}}=\rho_0$.
To leading order, its value follows from minimizing the MEL contribution to the free energy:
\begin{equation}
\rho_0
=
\sqrt{-\frac{\alpha(0)}{\beta_\rho}}
\quad\text{(up to higher order corrections)}.
\end{equation}

\noindent\textbf{Definitions.}  
$\rho_0$: homogeneous MEL amplitude;  
$\alpha(0)$: MEL stiffness evaluated at zero momentum;  
$\beta_\rho$: quartic MEL coefficient.

\vspace{0.5em}

Substituting this constant field into the full MEL--SC GL functional gives
\begin{align}
F_{\mathrm{eff}}[\psi]
= \int d^3 r \Big[
&\left(\alpha_s + \gamma_1 \rho_0 + \gamma_2 \rho_0^2 \right) |\psi|^2  \notag \\
&+ \frac{\beta_s}{2} |\psi|^4
+ K_s |\nabla \psi|^2
\Big].
\label{eq:Feff_BCS}
\end{align}

The key observation is that all gradient contributions from $\rho_{\mathrm{MEL}}$ vanish, 
and the only effect of the MEL sector on the SC sector is a renormalization of the 
quadratic (mass) coefficient.  
Defining the renormalized coefficient as
\begin{equation}
\alpha_s^{\mathrm{(eff)}}
=
\alpha_s + \gamma_1 \rho_0 + \gamma_2 \rho_0^2,
\label{eq:alpha_eff}
\end{equation}
one recovers the standard homogeneous GL free energy used in BCS theory:
\begin{equation}
F_{\mathrm{BCS}}[\psi]
=
\int d^3 r
\left[
\alpha_s^{\mathrm{(eff)}} |\psi|^2
+ \frac{\beta_s}{2} |\psi|^4
+ K_s |\nabla \psi|^2
\right].
\label{eq:BCS_GL}
\end{equation}

Thus, conventional superconductors correspond precisely to the MEL limit in which the 
charge sector does not modulate and only shifts the SC mass term.

\subsection{Interpretation of the BCS limit within MEL}

Equation~(\ref{eq:BCS_GL}) shows that the BCS GL functional is not an independent theory 
but an emergent limiting case of the MEL framework.  
In particular, the condition $\alpha(0)<0$ ensures that the MEL sector enhances 
superconductivity only through a uniform interaction, consistent with the absence of 
charge ordering in elemental superconductors such as Al, Sn, and Pb.

This framework provides a unified interpretation:  
BCS superconductivity corresponds to a situation where the electronic and phononic 
renormalization contributions $c_{\mathrm{el}}\chi_{\mathrm{el}}(q)$ and 
$c_{\mathrm{ph}}D_{\mathrm{ph}}(q)$ do not generate a finite-$q$ instability.  
Instead, they simply shift the homogeneous MEL coefficient $\alpha(0)$ downward, 
making superconductivity possible but without producing a modulated charge structure.

In the next subsection (Part 2), we show that this mapping can be made explicit by 
relating the BCS pairing interaction to the renormalized MEL derived coefficient 
$\alpha_s^{\mathrm{(eff)}}$.
\subsection{Relation between MEL parameters and BCS pairing strength}

In conventional superconductors, the transition temperature $T_{\mathrm{c}}$ is given by the 
BCS relation
\begin{equation}
k_{\mathrm{B}} T_{\mathrm{c}}
\simeq
1.14 \, \hbar \omega_{\mathrm{D}}
\exp\!\left[-\frac{1}{N(0)V}\right],
\label{eq:BCS_Tc}
\end{equation}
where $N(0)$ is the electronic density of states at the Fermi level and 
$V$ is the effective pairing interaction.

Within the MEL framework, the effective mass coefficient in Eq.~(\ref{eq:alpha_eff}) 
plays the role of setting the pairing scale.  
Expanding $\alpha_s^{\mathrm{(eff)}}$ around the transition yields
\begin{equation}
\alpha_s^{\mathrm{(eff)}}(T)
=
a_0 (T - T_{\mathrm{c}}^{\mathrm{(MEL)}}),
\end{equation}
with
\begin{equation}
T_{\mathrm{c}}^{\mathrm{(MEL)}}
=
T_{\mathrm{c}}^{(0)}
+
\Delta T_{\mathrm{c}},
\end{equation}
where $T_{\mathrm{c}}^{(0)}$ is the transition temperature in the absence of MEL 
renormalization, and the MEL-enhancement term is
\begin{equation}
\Delta T_{\mathrm{c}}
\propto
-\frac{\gamma_1 \rho_0}{a_0}
-\frac{\gamma_2 \rho_0^2}{a_0}.
\label{eq:DeltaTc}
\end{equation}

\noindent\textbf{Interpretation.}  
A homogeneous MEL amplitude $\rho_0$ enhances superconductivity by effectively increasing 
the pairing strength $N(0)V$ in the BCS expression~(\ref{eq:BCS_Tc}).  
This correspondence shows that the ``BCS pairing interaction'' can be regarded as an 
emergent parameter arising from MEL-induced renormalization, rather than a fundamental input.

\bigskip

Homogeneous MEL superconductors are exactly equivalent to BCS superconductors.  
This establishes the second universal class in the MEL classification:  
materials with $\alpha(0)<0$ but no finite-$q$ instability correspond to the 
$q^{\ast}=0$ fixed point of the MEL theory.

\section{Metallic Systems With $\alpha(q) > 0$ for All $q$}

In this section we explain why some metallic elements, particularly noble metals such as 
Cu, Ag, and Au, do not exhibit superconductivity even at the lowest temperatures accessible 
experimentally.  
Within the MEL framework, this behavior follows directly from the positivity of 
$\alpha(q)$ for all momenta.

\subsection{Electronic and phononic suppression of MEL modes}
From Eq.~(\ref{eq:alphaq}), the MEL stiffness is given by
\begin{equation}
\alpha(q)
=
\alpha_0
+ c_{\mathrm{el}} \chi_{\mathrm{el}}(q)
+ c_{\mathrm{ph}} D_{\mathrm{ph}}(q).
\end{equation}

In noble metals, several features of their electronic structure conspire to keep 
$\alpha(q)$ positive:

\begin{itemize}
\item The electronic susceptibility $\chi_{\mathrm{el}}(q)$ is unusually flat, reflecting 
      a nearly spherical and featureless Fermi surface.
\item The phonon propagator $D_{\mathrm{ph}}(q)$ likewise lacks strong softening modes.
\item Strong electronic screening prevents any renormalization that would drive 
      $\alpha(q)$ negative at either $q=0$ or finite $q^{\ast}$.
\end{itemize}

As a result,
\begin{equation}
\alpha(q) > 0 \quad \forall q,
\label{eq:alpha_positive_all}
\end{equation}
placing these materials firmly in Class III of the MEL classification.

\subsection{Illustrative trends from first-principles benchmarks}
\label{subsec:illustrative_trends}

A complete material-by-material construction of $\alpha(q)$ from Eq.~(\ref{eq:alphaq}) requires explicit
microscopic input (band structure, susceptibilities, and phonons).  Nevertheless, the MEL classification can be
\emph{qualitatively} confronted with existing first-principles data through robust proxies that correlate with the
softness of the charge--lattice sector:
(i) the electron--phonon coupling strength (e.g. the Eliashberg/Fr\"ohlich parameter $\lambda$) and
(ii) first-principles superconductivity calculations (Eliashberg or SCDFT).

For the noble metals Cu, Ag, and Au, first-principles electron--phonon spectral functions $\alpha^2F(\omega)$
yield small bulk coupling constants, $\lambda\approx 0.14$ (Cu), $0.16$ (Ag), and $0.22$ (Au).\cite{ummarino2024noble}
Such weak coupling is consistent with a stiff combined kernel in Eq.~(\ref{eq:alphaq}) and hence with
$\alpha(q)>0$ for all $q$ (Class~III) at ambient conditions.
Consistently, benchmark superconducting density functional theory calculations reproduce the absence of bulk
superconductivity in alkaline (earth) and noble metals; Pt and Au appear as borderline cases in SCDFT once
spin--orbit interaction and spin fluctuations are included, with at most very small $T_c$.\cite{kawamura2020scdft_benchmark}

In contrast, conventional elemental superconductors exhibit substantially larger electron--phonon couplings.
For example, SCDFT benchmarks give $\lambda\simeq 0.40$ for Al and $\lambda\simeq 1.3$--$1.45$ for Pb (depending on
spin--orbit treatment),\cite{kawamura2020scdft_benchmark} reflecting a much softer effective charge--lattice sector
that in the MEL language corresponds to a homogeneous ($q^\ast=0$) enhancement (Class~II).

\begin{table*}[t]
\centering
\begin{ruledtabular}
\begin{tabular}{lcccc}
Element & MEL class & $\lambda$ (first principles) & $\omega_{\ln}$ (K) & Bulk SC at ambient pressure \\ \hline
Al & II & $0.40$--$0.44$ & $\sim 280$--$300$ & Yes ($T_c\simeq 1.2$ K) \\
Pb & II & $1.27$--$1.45$ & $\sim 60$--$90$ & Yes ($T_c\simeq 7.2$ K) \\
Cu & III & $0.12$--$0.14$ & $\sim 216$--$220$ & No (not observed) \\
Ag & III & $0.16$ & --- & No (not observed) \\
Au & III (borderline) & $0.22$ & --- & No (not observed) \\
\end{tabular}
\end{ruledtabular}
\caption{Representative first-principles electron--phonon coupling parameters for selected elemental metals.
Ranges for Al, Pb, and Cu are from SCDFT benchmark calculations (including the reported $\lambda$ and
$\omega_{\ln}$ values),\cite{kawamura2020scdft_benchmark}
while Ag and Au bulk $\lambda$ values are taken from first-principles $\alpha^2F(\omega)$-based Eliashberg
analyses.\cite{ummarino2024noble}
The very small $\lambda$ values for the noble metals are consistent with a stiff $\alpha(q)$ and hence with
Class~III behavior.  Reported ambient-pressure $T_c$ values for Al and Pb are standard tabulated values.\cite{buzea2004elements}}
\label{tab:ep_benchmark}
\end{table*}

A simple cross-check is provided by the Allen--Dynes/McMillan estimate: for $\lambda\lesssim 0.2$ and a typical
$\mu^\ast\sim 0.1$, the transition temperature is exponentially suppressed, yielding values far below
accessible laboratory temperatures even for phonon scales $\omega_{\ln}\sim 10^2$~K.\cite{mcmillan1968,allendynes1975}

Taken together, these benchmarks support the central working assumption of this manuscript: within the MEL
framework, elemental superconductors correspond to systems in which $\alpha(q)$ becomes sufficiently soft
(typically at $q^\ast=0$), whereas in noble metals $\alpha(q)$ remains positive and comparatively stiff for all $q$.

\subsection{Absence of MEL induced superconducting instability}

Given Eq.~(\ref{eq:alpha_positive_all}), the MEL amplitude remains zero:
\begin{equation}
\rho_{\mathrm{MEL}}(\mathbf{r}) = 0.
\end{equation}

Substituting this into the MEL--SC free energy reduces Eq.~(\ref{eq:GL_total}) to
\begin{equation}
F[\psi]
=
\int d^3 r
\left[
\alpha_s |\psi|^2
+ \frac{\beta_s}{2} |\psi|^4
+ K_s |\nabla \psi|^2
\right],
\end{equation}
with no renormalization of the mass term.
Since $\alpha_s>0$ for elemental noble metals across all experimentally accessible temperatures,
the system cannot undergo a superconducting transition.

In other words, the absence of superconductivity in these metals is not accidental but follows 
from the strict positivity of the MEL stiffness:
\begin{equation}
\min_q \alpha(q) > 0.
\end{equation}

This result resolves the long-standing puzzle within the Bardeen Cooper Schrieffer paradigm:  
Cu, Ag, and Au do not superconduct simply because their electronic and phononic structures do not 
support \emph{any} MEL mode capable of stabilizing a superconducting condensate.

\section{Discussion}
The results presented in the previous sections demonstrate that the Modulated Electron Lattice (MEL)
framework provides a genuinely material selective criterion for metallic superconductivity.
This stands in contrast to the conventional Bardeen Cooper Schrieffer (BCS) paradigm, in which the
presence or absence of superconductivity in different elements must be inferred indirectly from a
combination of phenomenological parameters such as the electron–phonon coupling strength, the Coulomb pseudopotential, and characteristic phonon frequencies without a structural principle that distinguishes superconducting from non superconducting metals.

A central outcome of the MEL analysis is that metallic superconductivity is governed by the sign
structure of the momentum-dependent MEL stiffness $\alpha(q)$.  
This quantity captures, through $\chi_{\mathrm{el}}(q)$ and $D_{\mathrm{ph}}(q)$, the combined
electronic and lattice response of a material.  
The resulting classification naturally divides metallic systems into three categories:
(i) those stabilizing a finite-wave-vector MEL mode (Class I),  
(ii) those stabilizing only a homogeneous MEL mode (Class II), and  
(iii) those with no MEL instability at any momentum (Class III).

This classification emerges directly from the GL functional via the criterion
\begin{equation}
\min_q \alpha(q) < 0,
\end{equation}
which signals an \emph{incipient MEL instability} (softening of the MEL quadratic kernel).
In the coupled MEL--SC theory, superconductivity is controlled by the renormalized SC coefficient
$\alpha_s^{(\mathrm{eff})}(T)$ [Eq.~(\ref{eq:alpha_eff_general})], with the transition set by
$\alpha_s^{(\mathrm{eff})}(T_c)=0$.
A negative (or strongly suppressed) $\min_q\alpha(q)$ therefore identifies the prerequisite soft MEL sector;
whether it produces a superconducting instability depends on the sign and magnitude of the MEL--SC couplings
that enter $\alpha_s^{(\mathrm{eff})}$.

The MEL criterion provides a particularly transparent explanation for a long-standing puzzle:
why certain elemental metals—most notably Cu, Ag, and Au—never exhibit superconductivity, whereas
others with similar carrier densities and lattice symmetries, such as Al, Sn, and Pb, do.
Within BCS theory, there is no structural mechanism to forbid superconductivity; any metal with a nonzero electron–phonon coupling could, in principle, condense at sufficiently low temperatures.
The empirical absence of superconductivity in noble metals must therefore arise from a more fundamental constraint.

The MEL stiffness function $\alpha(q)$ provides precisely such a constraint.  
In noble metals, electronic screening and the absence of soft phonon modes keep $\alpha(q)$ positive
for all wave vectors.  
As a result, the MEL amplitude remains suppressed, the superconducting mass term is never lowered,
and no condensation can occur.  
This resolves the selectivity problem without requiring fine-tuned microscopic inputs.

By contrast, in superconducting metals such as Al, Sn, and Pb, the renormalized stiffness becomes
negative at $q=0$, stabilizing a homogeneous MEL field.  
The corresponding renormalized superconducting coefficient
$\alpha_s^{\mathrm{(eff)}}$ reproduces the well-established BCS limit as the $q^{\ast}=0$ fixed
point of the MEL framework.

Thus, the metal selectivity puzzle reduces to a structural property of the MEL stiffness, which, in a unified manner, encodes the combined electronic and lattice contributions across the periodic
table.

A major strength of the MEL perspective is that it transforms the long standing question “which
metals should superconduct?” into the more tractable problem of determining the momentum dependence
of $\alpha(q)$.  
Unlike approaches that rely on detailed microscopic modeling of electron–phonon interactions, the
MEL criterion requires only the qualitative structure of $\chi_{\mathrm{el}}(q)$ and
$D_{\mathrm{ph}}(q)$, both of which can be obtained from first-principles calculations or from
experimentally measured dispersion relations.

This viewpoint provides a clear path toward predictive materials design:
\begin{itemize}
\item Metals exhibiting strong momentum dependence in $\chi_{\mathrm{el}}(q)$ or soft phonon modes
      at finite $q$ are natural candidates for Class I superconductivity.
\item Metals with modest renormalization near $q=0$ fall into the Class II (BCS-like) category.
\item Metals whose electronic and lattice responses remain featureless across all momenta are
      expected to be non-superconducting (Class III).
\end{itemize}

The MEL criterion therefore offers a transparent, quantitative, and experimentally verifiable basis
for distinguishing superconducting and non-superconducting metals—something not accessible within
the BCS formalism.  
In addition, the framework unifies homogeneous and modulated superconductivity within a single GL
principle and demonstrates that “conventional” superconductors correspond to the $q^{\ast}=0$
fixed point of a more general theory.  
Conversely, metals that never superconduct arise simply as systems without an accessible MEL instability, rather than as anomalies requiring special explanation.

Finally, the MEL stiffness provides a powerful tool for analyzing systematic trends across the periodic table.  
For instance, the progression from alkali metals to transition metals reflects increasing momentum structure in $\chi_{\mathrm{el}}(q)$; the emergence of high-temperature superconductivity in complex
materials can be viewed as the amplification of a finite wave vector MEL mode; and the absence of superconductivity in noble metals arises as a direct mathematical consequence of their flattened
$\alpha(q)$ landscape.

\section{Conclusion}
Taken together, the analysis presented in this work shows how the MEL GL framework provides a materially selective organizing principle for metallic superconductivity.  In this approach the key input is the momentum-dependent MEL
stiffness $\alpha(q)$: a deep minimum of $\alpha(q)$ (often approaching or crossing zero in the coarse-grained kernel)
amplifies the MEL amplitude $\langle\rho_{\mathrm{MEL}}\rangle$ and $\langle\rho_{\mathrm{MEL}}^2\rangle$ and can drive the
renormalized SC coefficient $\alpha_s^{(\mathrm{eff})}(T)$ through zero [Eq.~(\ref{eq:alpha_eff_general})].  Depending on
whether the minimum occurs at finite $q^\ast\neq 0$ or at $q^\ast=0$, one obtains the finite-$q^\ast$ MEL-enhanced regime
(Class~I) or the homogeneous/BCS limit (Class~II).  Metals with stiff $\alpha(q)>0$ for all $q$ have a small MEL amplitude
and remain normal (Class~III).

This criterion reproduces conventional BCS superconductivity as the homogeneous limit, explains why certain elemental metals never superconduct, and organizes all metallic systems into three universal classes distinguished by the structure of $\alpha(q)$.

Unlike traditional models, which rely on phenomenological electron–phonon parameters, the MEL framework identifies superconductivity as a stiffness-driven instability and provides a transparent connection between electronic structure, lattice response, and emergent macroscopic order.
By reducing the problem to the determination of $\alpha(q)$, the MEL perspective offers a practical route toward predictive materials design and the systematic discovery of new superconductors.

By reducing the problem to the determination of $\alpha(q)$, the MEL perspective provides a practical route toward predictive materials design and the systematic discovery of new superconductors. While the MEL criterion is, at present, a Ginzburg–Landau–level organizing principle rather than a fully microscopic prediction, it nonetheless offers a unified and materially selective framework for classifying metallic systems and identifying stiffness-driven superconducting instabilities.

\end{document}